# The Aware Cricket Ground

**Wazir Zada Khan[1], Mohammed Y Aalsalem[2] and Quratul Ain Arshad[3]**

[1] School of Computer Science, University of Jazan
Jazan, PoBox # 114, Kingdom of Saudi Arabia

[2] School of Computer Science, University of Jazan
Jazan, PoBox # 114, Kingdom of Saudi Arabia

[3] School of Computer Science, University of Jazan
Jazan, PoBox # 114, Kingdom of Saudi Arabia

**Abstract**
The most profound technologies are those that disappear. They weave themselves into fabrics of everyday life until they are indistinguishable from it [1]. This research work is a mere effort for automated decision making during sports of most common interest leveraging ubiquitous computing. Primarily cricket has been selected for the first implementation of the idea. A positioning system is used for locating the objects moving in the field. Main objectives of the research are to help achieve the following goals. 1) Make Decisions where human eye can make error due to human limitations. 2) Simulate the Match activity during and after the game in a 3D computerized Graphics system. 3) Make various types of game and performance analysis of a certain team or a player.
**Keywords:** *Ubiquitous Computing, Decision making, Context Awareness, Sports*

## 1. Introduction

Everyone has liking and disliking and when it is the matter of National sports things get more sensitive and touchy. Any keen follower of cricket knows that when a batsman is dropped during an innings he is given a life. Whereas when he is declared `out' it is death for him. Wrong decisions at the time of a climax in any game can lead to ultimate disappointment and a state of disparity for its fans. Remember, a single wrong decision can spoil the entire match. Sports science divides coaching into activities that take place before, during and after competition [2]. Computer Science already supports a number of coaching activities before and after competitions, such as strategy development and performance evaluation. Computer-based, semi-automatic observational systems support coaches by combining quantitative statistics with qualitative video analysis [3]. A number of misjudged decisions by umpires due to incapability or simple carelessness led to totally spoiled matches. There are numerous examples which one can quote in this respect. "Test in England where an umpire gave three batsmen out on no-balls. TV replays showed that the bowlers had clearly violated the no-ball rule but the umpire never called.

A series of cricket matches (VB Series 2005) played between West Indies, Australia and Pakistan teams, at the end Australia won but what happened? "Pakistanis claimed umpires favor vocal Aussies". Many other examples are there where game is affected badly due to wrong decisions of umpires and these decisions are not only taken once but repeated many times cautiously.

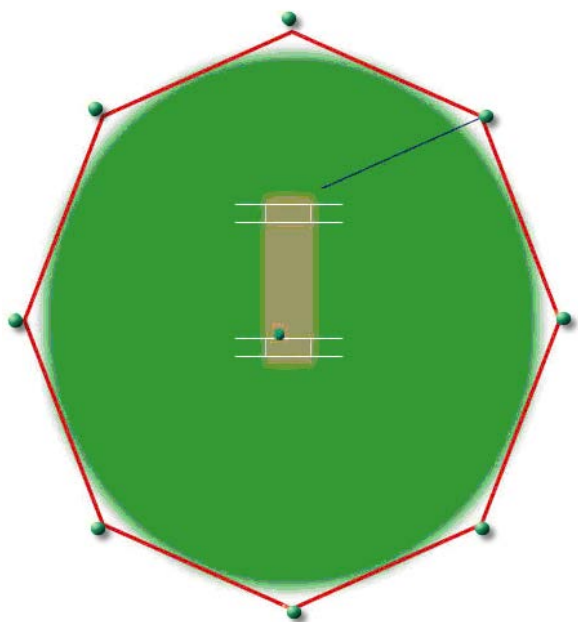

Fig 1: Placement of Access points

There exist a number of problems which cause dissatisfaction of the players and game lovers when a decision is of great importance and a misjudgment done by an umpire crumbles the whole situation. The Automated ball tracking system proposed in this paper helps solve a number of key issues in judgment and decision making during the game using ubiquitous computing. Problems addressed in this paper are:

A. No Ball Decision

B. Fielders in the Ring in First 15 Overs

C. Leg Before Wicket Decision (Trajectory Calculation)

A number of other useful information can also be extracted from the data collected by the system.
- Bowling Speed
- Fielding Analysis
- Striking Analysis
- Strike Power etc.

The structure of our paper is organized as follows:
Section 2 reviews the related literature. Section 3 presents our proposed system. In Section 4 our proposed system is compared with the existing system and Section 5 concludes the paper.

## 2. Related Work

A very little work is done in this area, thus new tools and systems are required for supporting all kinds of human activities. The systems must have the goals of accuracy, robustness and Security. A brief detail of the existing work is described below:

M. Beetz et al [4, 5, 6] have developed the Football Interaction and Process Model FIPM and a software system that can acquire, interpret, and analyze this model. The system can acquire models of player skills, infer action-selection criteria, and determine player and team strengths and weaknesses. This system extracts the events of the game based only on the position of players and the ball. It has been implemented using the microwave sensors mechanism for automatic analysis of the football during the game. The FIPM game analysis system assumes a real-time positioning system that can continually track the positions of the feet of all players and the ball with an accuracy of a few centimeters and provides position data for the ball that enables the game analysis system to accurately detect ball contacts of the players. They have tested it on the RoboCup simulation league and have received promising results.

Jonathan Hey and Scott Carter [7] have designed and implemented an enhanced table tennis practical table called Memory Tennis which projects a lasting image of the last place the ball struck on the vertical practice wall. Using this, the players can determine their accuracy immediately and actively compensate for poor shots. Additionally, players can get statistics on past performance. The system can project targets for players to hit and act as coaching program as player's performance improves.

A. G. Foina et al [8] have designed a tool to help sports coaches to analyze their players using an RFID technology connected to 3-layer software. The system has two tracking modes, one for 2D player location in the field and a 3D mode to capture player's movement in small area sports. It has two report modes also, one for real time report displaying each player's actual location in the field and the other to present reports with statistics of the player. The system described by authors is neither fully implemented nor tested in a true sports competition.

Zoe Drey and Charles Consel [9] have presented a visual programming language called Pantagruel which is end-user oriented. Their approach is open ended in that Pantagruel integrates a language to describe a ubiquitous computing environment. To facilitate the programming of the orchestration logic, they have developed a visual tool that uses a sensor-controller-actuator paradigm. An Orchestration logic collects context data from sensors, combines them with a controller, and reacts by triggering actuators. They have assessed the usability of this paradigm with a successful user study conducted with 18 non-programmer participants. Furthermore, their visual programming environment offers the developer an interface that is customized with respect to the environment description. The authors have implemented a visual environment to prototyping ubiquitous computing applications. They have also developed a compiler for Pantagruel that targets a domain-specific middleware.

Ed H. Chi et al [10] have propelled ubiquitous computing into the extremely hostile environment of the sparring ring of a martial art competition. The authors have described an ovel wearable computing system called TrueScore TM SensorHogu for supporting an extreme human activity namely Taekwondo competition sparring. It uses piezoelectric sensors to sense the amount of force that has been delivered to a competitor's body protector and wirelessly transmits this signal to a computer that scores and displays the point. Their objective is to support the judges in scoring the sparring matches accurately while preserving the goals of merging and blending into the background of the activity.

Ramy Asselin et al [11] have presented the design and evaluation of their proposed system called Personal Wellness Coach System (PWC). It supports a range of activities including health data collection, interpretation, feedback and self monitoring. They have implemented a wearable computing platform which consists of an array of sensors and application software to motivate users to reach fitness goals and prevent harm in a real time environment. Evaluation and testing has been done on 65 users in four groups and the results indicate that this system is an effective tool that is less intrusive and helps to improve user efficiency.

## 3. Proposed System

This paper focuses on finding the location of the ball during the game at any given instance and thus making the desired calculations whenever needed for a certain analysis. The System makes decision making it a very easy and painless task. Along-with this it makes it possible to animate the whole game at any time in a 3D Computer animation. Being a high precision system and a bulk data producer, it can be used to make a vast variety of decisions based on this information.

3.1 Working of Proposed System:

Sensors are placed at various hotspots in the playground, cricket ball, and in the shoes of the players. With the help of these sensors information about all aspects of the game is collected and sent to a central computer for further processing. The Computer collects the data, formats it for storage and stores it in a ways that it can be used for any purpose in the future. Data received from various sensors placed in the moving objects is processed and we get the distance of object from certain predefined point. Using the same mechanism we address the problems in the order as mentioned above.

3.1.1 No Ball Decision:

We only address the issue of over stepping the popping crease by the bowler. To detect whether a specific delivery is valid or is a no ball. We use a triangular placement of sensors as shown in Fig- 2.

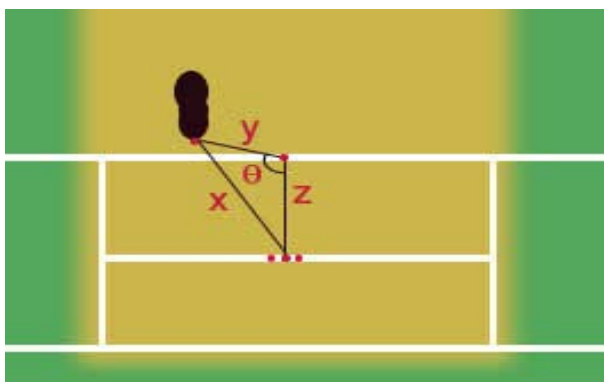

Fig 2: Calculating no Ball

We achieve a triangular assembly of sensors by mounting two sensors statically in the crease and the third one in the shoe of the bowler. Sensors in all the three positions talk to the central computer and communicate their (x, y) coordinates. Once we have the coordinates of the sensors we can calculate the length of the sides of the triangle using distance formula as given in equation-1:

$$d = \sqrt{(x_2 - x_1)^2 + (y_2 - y_1)^2}$$

Eq 1: Distance Formula

The distance formula provides the lengths of all the three sides of the triangle and now we can calculate the angles involved in the creation of the triangle using the formula given in the equation-2. Here a check is made before calculating the angle, as we know that the length of the side which is labeled z, now if the side x gets smaller in length then the side z it give an automatic judgment of not being a no ball. If the angle θ is greater than 90° then it is assumed that the bowler has over stepped the popping crease line, indicating a no ball. Once a no ball is detected the system alarm for no ball and communicates it to the umpire as well as to the scoring system.

$$x^2 = y^2 + z^2 + 2yzCos\Theta$$

Eq 2: Cosine formula

3.1.2 Fielders in the Ring in First 15 Overs:

As we know that sensors have already been mounted in the shoes of all the players and they are in constant contact with the central computer, so it is quit easy to calculate the number of fielders in any location of the cricket ground.

3.1.3 Leg-Before Wicket Decision (Trajectory Calculation):

Trajectory calculation for Leg-Before Wicket so far has been achieved through the use of Hawk eye camera and other similar image processing and computer vision technique using cameras mounted for capturing the match events. We are deploying a different mechanism for calculation of the ball trajectory. We have a special type of sensor which responds to the nearby microwave radar, as the moving ball communicates its position to the receptor once every 100th of a second. So we have quite enough amounts of data about the movement of the ball, which can be used to calculate the parabolic curve of the ball traveling across the wicket.

Trajectory calculation can be precisely done on the basis of points obtained from the previous motion of the ball in the prescribed plane.

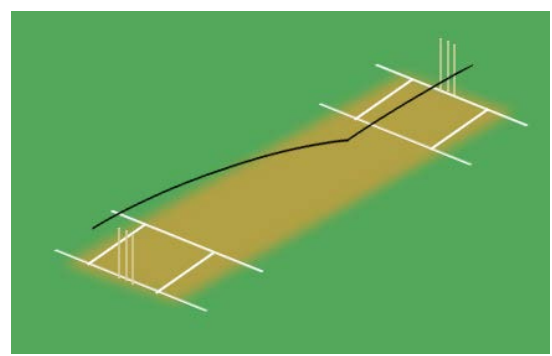

Fig 3: Trajectory of the Ball

$$R^2 = D^2 + (R - Yd)^2$$

Now let's do a little algebra:

$$R^2 - D^2 = (R - Yd)^2$$

$$sqrt(R^2 - D^2) = R - Yd$$

$$Yd = R - sqrt(R^2 - D^2) \quad [5]$$

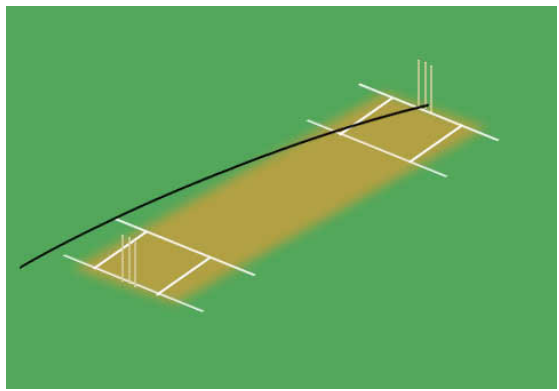

Fig 4: Trajectory of the Ball

### 3.2 Game Analysis:

Analysis of various activities of players during game can be monitored through their movements and a strategic plan can also be setup. Bowling analysis of the game can be performed automatically and conclusions can be made. Graphs of the strike rate, strike power will also be created.

### 3.3 Proposed System Architecture:

Our system is organized as a four-tier layered architecture according to their functionality, which are given as under.

#### 3.3.1 Data Generation/ Data Gathering Layer:

Sensor/ hardware (camera's etc) will sense/ capture the event and generate the data.

#### 3.3.2 Data Communication Layer:

The generated data is collected from sensors and other different Hardware (i.e. camera's etc) through the network of an interconnected access points and cables and is transferred to the Main Software processing Unit.

#### 3.3.3 Data Processing Layer:

The main software will process the collected data

#### 3.3.4 Data Presentation/Data Reporting Layer:

Processed Data will be used/ presented for further actions according to user request like:

- For umpires in making a decision.
- For coaches to analyze the game and future planning.
- For viewers in making analysis & statistics.

The stored data can be used for future analysis by the players and team management to improve their game, overcoming their weaknesses. It also finds out the weakness of other players.

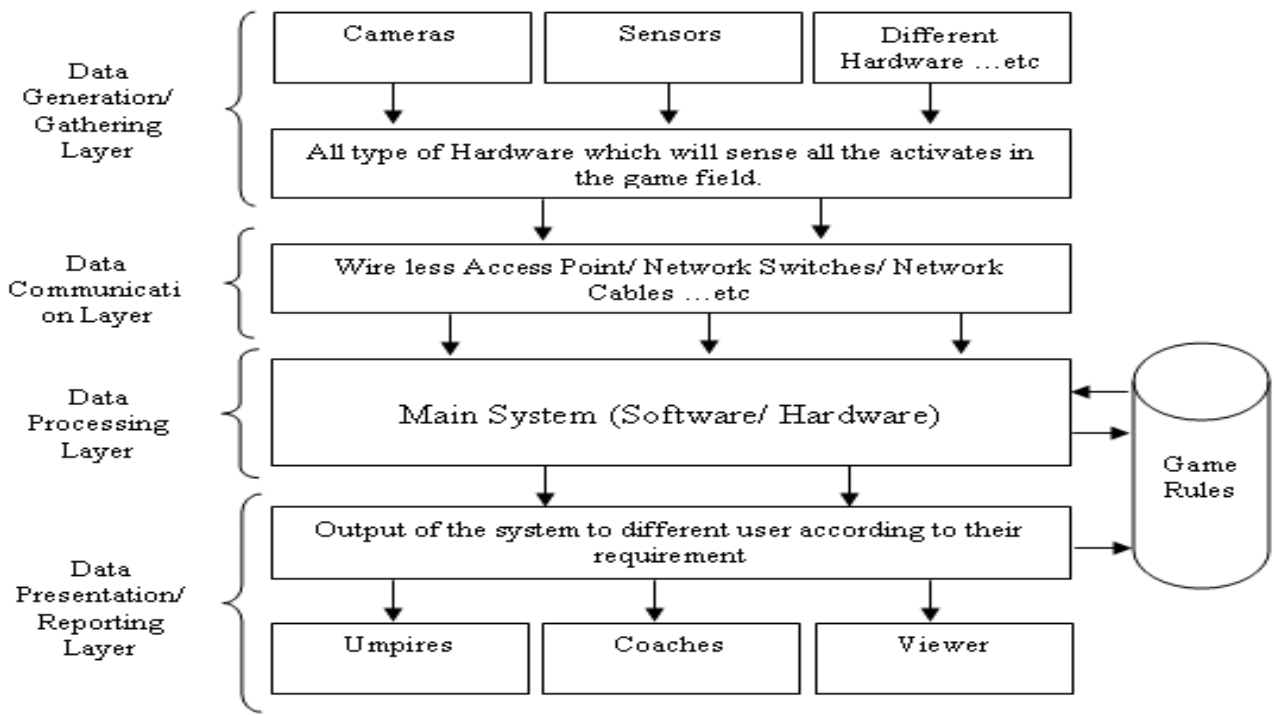

Fig.5 Proposed System Architecture

## 4. Comparison With Existing Systems

Currently for decision taking according to the cricket rules, umpire's decision is final but however if there is any doubt, the ground umpire can take help from the third umpire who uses recorded videos for giving his final decisions. Many softwares are available which can calculate the trajectory of the ball. Hardware like speed calculation cameras are available which can calculate the speed of the ball. Factors like wind speed, wind directions, humidity in air, pitch condition, ball's conditions can determine the direction of the ball. Some software can also help in determining statistical analysis and scoring the game.

All the available techniques for checking different activities of cricket work independently. Comparatively, our proposed system works in a combined fashion assisting in correct decision making. It considers all the important activities of the game which if not considered during decision making may spoil the whole game.

## 5. Conclusion & Future Work

A work in progress of a tool designed to help game umpires, sports coaches to help in decision making, analyzing the players using pervasive computing connected to a four layer system is described in this paper. An indoor location tracking system for pervasive and sensor-based computing environments is developed by MIT called Cricket [12]. Cricket provides fine-grained location information space identifiers, position coordinates, and orientation to applications running on handhelds, laptops, and sensor nodes. In the same way, systems for tracking activities of outdoor games should also be implemented. Regarding the game of cricket, future challenges include tracking the ball in real-time, sensing the no ball etc Accuracy and in-time detection of problems of the outdoor games is also a big challenge for the researchers.

**Acknowledgment**

The authors wish to acknowledge the anonymous reviewers for valuable comments.

**Wazir Zada Khan** is currently with School of Computer Science, Jazan University, Kingdom of Saudi Arabia. He received his MS in Computer Science from Comsats Institute of Information Technology, Pakistan. His research interests include network and system security, sensor networks, wireless and ad hoc networks. His subjects of interest include Sensor Networks, Wireless Networks, Network Security and D igital Image Processing, Computer Vision.

**Dr. Muhammad Y Aalsalem** is currently dean of e-learning and assistant professor at School of Computer Science, Jazan University. Kingdom of Saudi Arabia. He received his PhD in Computer Science from Sydney University. His research interests include real time communication, network security, distributed systems, and wireless systems. In particular, he is currently leading in a r esearch group developing flood warning system using real time sensors. He is Program Committee of the International Conference on Computer Applications in Industry and Engineering, CAINE2011. He is regular reviewer for many international journals such as King Saud University Journal (CCIS-KSU Journal).

**Quratul Ain Arshad** received her BS in Computer Science from Comsats Institute of Information Technology, Pakistan. Her research interests include network security, trust and reputation in sensor networks and ad hoc networks. Her subjects of interest include Network Security, Digital Image Processing and Computer Vision.